\documentclass[12pt]{article}
\usepackage{amsfonts,amssymb,latexsym,amscd}
\usepackage{graphicx,epsfig}
\title{\textbf{Evolution of shifted cosmological parameter and  shifted dust matter in a two-phase tachyonic  field universe}}
\date{}
\author{Murli Manohar Verma\footnote{sunilmmv@yahoo.com}\quad and \quad Shankar Dayal Pathak\footnote{prince.pathak19@gmail.com}\\ \emph{Department of Physics} \\\emph{ Lucknow University, Lucknow  226 007, India }}
\begin{document}
\maketitle

\begin{abstract}
We  propose  a model of the evolution of the  tachyonic  scalar field over two phases in the universe. The  field components do not interact in phase I,  while in the subsequent phase II, they change flavours due to relative suppression of the radiation contribution. In  phase II, we allow them  to interact mutually  with time-independent perturbation in their  equations  of state,  as  Shifted Cosmological Parameter (SCP) and  Shifted Dust Matter (SDM).   We determine the solutions of their scaling  with the cosmic redshift in both phases. We further suggest   the  normalized Hubble function  diagnostic, which,  together with the low- and high-redshift  $H(z)$  data and the concordance values of the present density parameters from the CMBR, BAO statistics etc.,  constrains  the strength of interaction,  by imposing the viable conditions to break degeneracy in 3-parameter $(\gamma, \varepsilon, \dot{\phi}^2)$ space. The range of redshifts $(z=0.1$  to $z=1.75)$ is chosen to highlight the role of interaction during structure formation,  and  it may lead to  a future analysis of power spectrum in this  model \emph{vis a vis}  Warm Dark Matter (WDM)  or  $\Lambda$CDM  models.  We further calculate  the influence of interaction in determining the age of the universe at the present epoch,  within the degeneracy space  of  model parameters.

\end{abstract}

\section{\textbf{Introduction}} The observed accelerated expansion of the universe \cite{x1,x2,x3} is thought to be  driven by some exotic field, called dark energy,  with an equation of state (EOS) very close to $w_{\lambda}=-1$ at the present epoch \cite{x4,x5,x6}.
A class of scalar fields  is one of the promising candidates of dark energy \cite{k1,i1,j1,k2,k3,k4}.  Among itself,  tachyonic scalar field arising from string theory \cite{n1} (for different reasons though) appears more relevant than the conventional non-relativistic scalar field in form of quintessence, and it is widely used in literature \cite{l1,m1,o1,y0}. One of its  reasons  is that the Lagrangian adopted in tachyonic scalar field is relativistic  which is more profound and appealing than its non-relativistic counterpart.

In this paper,  we propose  a model of  the  evolution of the tachyonic field  universe  over  two phases--- namely,  non-interacting (phase I) and interacting (phase II)---,   respectively.  The general evolution, however, may span several such phases with different intervals  of time distinguishing each other in terms of their respective interaction strengths.  In Sec. 2 of  our present work,  we investigate  phase I, when  the tachyonic scalar field has two components--- one is radiation while  the other is an  unknown stuff.  This later component  has  negative pressure and thus  mimics the  cosmological constant that may have caused the  inflation  early on  during  this  phase.  It is further decomposed into  two components--- true cosmological constant and  matter with negative energy density but zero EOS.

 The same tachyonic scalar field manifests itself in form of  two components in  phase II.  If we take one component as pressureless dark matter with $w_{m}=0$,  then other one is found to behave as the cosmological constant with $w_{\lambda}=-1$. The dark matter component includes the baryonic contributions having the same equation of state while any contributions arising from radiation would be almost negligible in a matter dominated universe. Several workers have investigated the cosmological behaviour of   tachyonic scalar field having this composition \cite{l1,m1,o1,y0}.

  In Sec. 3  we study the behaviour of  two new components.  One component of tachyonic scalar field is cosmological constant with  $w$ slightly perturbed  from $-1$. This component is called  Shifted Cosmological Parameter (SCP) with $w=-1+\varepsilon(t)$, where $\varepsilon(t)$ is a small perturbation introduced in the equation of state(EOS) of the true cosmological constant. The other component is dark matter with $w\simeq 0$.  Since it is not  exactly a  pressureless dust, we call it  Shifted Dust Matter (SDM).  Given the form  of energy density and  EOS of the SCP  we find its pressure. This further leads to the pressure and  EOS of SDM.

The non-zero pressure of SDM   contributes to the total kinetic energy of dark matter particles,  and justifiably raises the status of cold dark matter to warm dark matter(WDM). This connection is also supported by the current work in favour of  eV- (or  perhaps  keV-)  mass sterile neutrino\cite{a1} that elevates the energy of the cold dark matter,  or  of   WDM \cite{a2}. It is clearly seen that the perturbation introduced in EOS of  SCP influences the EOS of the dark matter  without involving any change in the nature of overall tachyonic scalar field.
We further study the interaction among SCP and SDM  which plays a major role  in the post-recombination era in structure formation imprints on the power spectra of  WDM \cite{a2},   or Lambda Cold Dark Matter ($\Lambda$CDM) \cite{a3}   models. Interaction is further important around the present epoch in sharply  reducing the value of SCP energy density, which suggests a possible solution to the cosmological constant problem, namely, why the present value of cosmological constant is merely $\approx10^{-123}$ of its value in  very early universe.  This interaction emerges as a natural corollary of the effective predominance of dark energy through a long evolution of the universe wherein it seems hard to believe  that this component, though drove  the dynamics of the universe by accelerating it, nevertheless failed to interact with another major co-existing component in form of dark matter.

In Sec. 4, we introduce a normalized Hubble function  $E(x)$ as a diagnostic to constrain the interaction strength in degenracy space of the cosmological parameters in this model.  We use the data of the Hubble parameter at different redshifts ($z=0.1$ to $z=1.75$) and the present concordance values of density parameters  from  Cosmic Microwave Background Radiation (CMBR) and Baryon Acoustic Oscillations (BAO) data sets in a spatially flat universe. We also discuss the cosmic age issue in Sec. 5 to highlight the role of interaction in determining it.

\section{\textbf{Phase I--- Non-Interacting era---radiation  and  shifted cosmological parameter }}
This phase consists of the evolution of two main tachyonic field components in the universe with further  resolution into three components.  The complete action  given by (taking $c=1$ in this paper)
\begin{eqnarray}\mathcal{A}=\int d^{4}x \sqrt{-g} \left(\frac{R}{16\pi G}- V(\phi)\sqrt{1-\partial^{i}\phi\partial_{i}\phi}\right)\label{n1}\end{eqnarray}
 couples gravity to tachyon  scalar field.  Here,  $V(\phi)$ is potential of the field which can be determined for an  \emph{a priory}  form of evolution of the scale factor, such as the quasi-exponential expansion \cite{y0}.  The corresponding  field Lagrangian  is given as
\begin{eqnarray}\mathcal{L} (\phi, \partial^i\phi)=-V(\phi)\sqrt{1-\partial^{i}\phi\partial_{i}\phi}\label{n2}.\end{eqnarray}  The energy-momentum stress tensor for the  field defined as \begin{eqnarray}T^{ik}=\frac{\partial\mathcal{L}(\phi, \partial^i\phi)}{\partial(\partial_{i}\phi)}\partial^{k}\phi-g^{ik}\mathcal{L}(\phi, \partial^i\phi)\label{n3}\end{eqnarray} yields energy density and  pressure,  respectively,  as
\begin{eqnarray}\rho(\phi)=\frac{V(\phi)}{\sqrt{1-\partial^{i}\phi\partial_{i}\phi}}\label{n5}\end{eqnarray} and

\begin{eqnarray}P(\phi)=- V(\phi)\sqrt{1-\partial^{i}\phi\partial_{i}\phi}\label{n4}.\end{eqnarray}

For spatially homogeneous tachyonic scalar field,  (\ref{n5}) and (\ref{n4}),  respectively,  become
\begin{eqnarray}\rho(\phi)=\frac{V(\phi)}{\sqrt{1-\dot{\phi}^{2}}}; \qquad P(\phi)=-V(\phi)\sqrt{1-\dot{\phi}^{2}}\label{m1}\end{eqnarray}  where, and henceforth in this paper, an overdot implies the derivative with respect to time.

We break  the energy density and pressure of tachyonic scalar field into  two components $A$ and  $B$,   which do  not interact mutually. Thus, $P=P _{A}+ P_{B}$  and  $\rho=\rho_{A} + \rho_{B} $.  The first term in pressure is $P_{A}= \frac{\dot{\phi}^{2}V(\phi)}{\sqrt{1-\dot{\phi}^{2}}}$ and we take it  as the radiation pressure. Then energy density of this component  (radiation)  is $\rho_{A}=\frac{3\dot{\phi}^{2}V(\phi)}{\sqrt{1-\dot{\phi}^{2}}}$  with EOS  $w_{r}=1/3$.
Next,  for the second component,  we have
\begin{eqnarray}\rho_{B}= \frac{(1-3\dot{\phi}^{2})V(\phi)}{\sqrt{1-\dot{\phi}^{2}}}; \qquad
P _{B}=-\frac{V(\phi)}{\sqrt{1-\dot{\phi}^{2}}}\label{m3}\end{eqnarray} with its  EOS $w_{B}=-1/(1-3\dot{\phi}^{2})$.

It can be seen that as  $\dot{\phi}^{2}\rightarrow 0$,   $w_{B} \rightarrow -1$,   and therefore,  we call  the component $B$  as `Shifted Cosmological Parameter' (SCP) with perturbed EOS   undergoing a shift  $\eta=-3\dot{\phi}^{2}/(1-3\dot{\phi}^{2})$.

 The SCP  gets further decomposed  into two components  $B_{1}$ and $B_{2}$.  Hence,  $\rho_{B}=\rho_{B_{1}} + \rho_{B_{2}}$ with $\rho_{B_{1}}=\frac{V(\phi)}{\sqrt{1-\dot{\phi}^{2}}}$ and   $\rho_{B_{2}}=\frac{-3\dot{\phi}^{2}V(\phi)}{\sqrt{1-\dot{\phi}^{2}}}$.  The corresponding pressure components  are  given as $P_{B_{1}}=\frac{-V(\phi)}{\sqrt{1-\dot{\phi}^{2}}}$  and   $P_{B_{2}}=0 $,  with total SCP pressure  $P_{B}=P_{B_{1}} + P_{B_{2}}$. Since the EOS of component $B_{1}$ is $w_{B_{1}}=-1$,  therefore,  we recognize it as the true cosmological constant  with $\rho_{\lambda}=\frac{V(\phi)}{\sqrt{1-\dot{\phi}^{2}}}$ and $P_{\lambda}=\frac{-V(\phi)}{\sqrt{1-\dot{\phi}^{2}}}$. The second component $B_{2}$ is a pressure-less  exotic matter with $w_{B_{2}}=0$ and  negative energy density   $\rho_{B_{2}}=\rho_{m}=\frac{-3\dot{\phi}^{2}V(\phi)}{\sqrt{1-\dot{\phi}^{2}}}$.

In  the absence of interaction,  the laws of conservation  of energy for each component--- radiation,  cosmological constant and exotic matter--- are given,  respectively, as

\begin{eqnarray}\dot{\rho_{r}} + 3H(1 + w_{r})\rho_{r} = 0\label{m4}\end{eqnarray}
\begin{eqnarray}\dot{\rho_{\lambda}} + 3H(1 + w_{\lambda})\rho_{\lambda} = 0\label{m5}\end{eqnarray}
\begin{eqnarray}\dot{\rho_{m}} + 3H(1 + w_{m})\rho_{m} = 0\label{k1}\end{eqnarray}
where $H(t)= \dot{a}(t)/a(t)$  is the Hubble parameter with $a(t)$  as the scale factor at some epoch $t$ lying in this phase. Using redshift $z$ scaling as $1+z=a_0/a(t)$  with  the present scale factor $a_0$, the solution of (\ref{m4}) with $w_{r}=1/3$ is obtained as
\begin{eqnarray}\rho_{r}=\rho^{i}_{r}\left(\frac{1+z}{1+z_{i}}\right)^{4}\label{m6}\end{eqnarray}
where $\rho^{i}_{r}$ and $z_i(\leq z)$  are   the energy density of radiation  and  redshift, respectively,  at the  epoch $t_i$  when   matter begins to interact with cosmological constant and the next phase starts.

Similarly, from (\ref{m5})
\begin{eqnarray}\rho_{\lambda}=\rho^{i}_{\lambda}=\mbox{constant} \label{k2}\end{eqnarray}
and from (\ref{k1})  we have   $\rho_{m}$,   with $w_{m}= 0 $,  scaling  as
\begin{eqnarray}\rho_{m}=\rho^{i}_{m}\left(\frac{1+z}{1+z_{i}}\right)^{3}\label{m7}.\end{eqnarray}

In phase I,  matter does not play  effective role in the evolution of universe owing  to the dominant counterpart of radiation. Thus,  the Friedmann equation for spatially flat ($k=0$) universe in the non-interacting phase is given as

\begin{eqnarray}H^{2} = \left(\frac{\dot{a}}{a}\right)^{2} = \frac{8\pi G}{3}\left[\rho_{r}+\rho_{m}+\rho_{\lambda}\right]. \label{m8} \end{eqnarray} In terms of (\ref{m6}), (\ref{k2}) and  (\ref{m7}), it changes to the form

\begin{eqnarray}H^{2}(z) = H^{2}_{i}\left[\Omega_{r}^{i}\left(\frac{1+z}{1+z_{i}}\right)^{4}+\Omega_{m}^{i}\left(\frac{1+z}{1+z_{i}}\right)^{3}+\Omega_{\lambda}^{i}\right] \label{m10} \end{eqnarray} where $H_i$, $\Omega^{i}_n (=\rho^i_n/\rho^{i}_{c})$  and $\rho^{i}_{c}(={3H^{2}_{i}}/{8\pi G})$ are,  respectively, the  Hubble parameter,  density parameter (with subscript $n$ denoting each component) and critical energy density at the  epoch  $z_{i}$.

\section{\textbf{Phase II--- Interacting era---shifted cosmological parameter and shifted dust matter}}

 By the end of the non-interacting era at $t=t_i$,  radiation loses  strength  and  ceases to   influence the large scale  dynamics of the universe appreciably. In the subsequent interacting era (phase II), thus,  the same tachyonic scalar field breaks  into two components  (cosmological constant and matter). The energy density and pressure of cosmological constant  are given as
\begin{eqnarray}\rho'_{\lambda}=V(\phi)\sqrt{1-\partial^{i}\phi\partial_{i}\phi}; \qquad P'_{\lambda}=-V(\phi)\sqrt{1-\partial^{i}\phi\partial_{i}\phi}\label{m11}\end{eqnarray} with $w'_{\lambda}=-1$,  (here and henceforth in this paper, a prime denotes the quantities pertaining to the interacting phase),   and for matter as
\begin{eqnarray}\rho'_{m}=\frac{V(\phi)\partial^{i}\phi\partial_{i}\phi}{\sqrt{1-\partial^{i}\phi\partial_{i}\phi}}; \qquad P'_{m}=0 \label{m12}\end{eqnarray} with $w'_{m}=0$.

If we introduce a  small perturbation $\varepsilon(t)$ in the EOS of cosmological constant,  the new EOS becomes  $w'_{\lambda}=-1+\varepsilon(t)$. Due to this change, the new  incarnation  of this component is called  the  shifted cosmological parameter(SCP). Although $\varepsilon(t)$ is a  function of time in general, yet  here,  we assume it to be a very small constant.  Considering  energy density of SCP, $\rho'_{\lambda}=V(\phi)\sqrt{1-\partial^{i}\phi\partial_{i}\phi}$, its  pressure can be  given as
\begin{eqnarray}P'_{\lambda}=-V(\phi)\sqrt{1-\partial^{i}\phi\partial_{i}\phi} + \varepsilon V(\phi)\sqrt{1-\partial^{i}\phi\partial_{i}\phi}\label{n8}.\end{eqnarray}  Thus,  pressure of the second component, namely, shifted dust matter (SDM) is \begin{eqnarray}P'_{m}=-\varepsilon V(\phi)\sqrt{1-\partial^{i}\phi\partial_{i}\phi}\label{n9}\end{eqnarray}  whose  EOS   now becomes  (with its  energy density given in  (\ref{m12}))
 \begin{eqnarray}w'_{m}= -\frac{\varepsilon}{\partial^{i}\phi\partial_{i}\phi}+\varepsilon\label{n10}\end{eqnarray}
while the EOS of  the  total  tachyonic scalar field is
 \begin{eqnarray}w_{tach}=(\partial^{i}\phi\partial_{i}\phi-1)\label{n11}.\end{eqnarray}
Here,  we  consider  spatially homogeneous scalar field, thus making  for SCP,  $\rho'_{\lambda}=V(\phi)\sqrt{1-\dot{\phi}^{2}}$ and $P'_{\lambda}=(\varepsilon-1)V(\phi)\sqrt{1-\dot{\phi}^{2}}$, while for SDM,  $\rho'_{m}=V(\phi)\dot{\phi^{2}}(t)(1-\dot{\phi^{2}})^{-1/2}$ and $P'_{m}=-\varepsilon V(\phi) (1-\dot{\phi}^{2})^{1/2}$ with $w'_{m}=-\varepsilon/\dot{\phi}^{2}+\varepsilon$.

In order to have negative pressure in the SCP component the perturbation condition $\varepsilon<1$ must be satisfied. This is because, if $\varepsilon\geq1$, dark energy will lose its  effective role in causing acceleration. Thus,  the following inequalities  must  hold with $\dot{\phi}^{2}<<1$.
\begin{enumerate}
\item[(i)]  If $0<\varepsilon<1$ then $P'_{m}<0$ with $w'_{m}<0$,  and  $P'_{\lambda}<0$ with $w'_{\lambda}<0$.

\item[(ii)] Alternatively, if $\varepsilon<0$ then $P'_{m}>0$ and $P'_{\lambda}<0$ (since $w'_{\lambda}<-1$ in this case, the  SCP behaves like phantom energy).
\end{enumerate}

In case (i) above, it can be seen that with $\varepsilon>0$  both SCP and SDM act gravitationally in the similar way, while in case (ii),  with $\varepsilon<0$,  their respective roles  in the dynamics of the universe are mutually opposite.

We have two simple  justifications  for calling upon the role of interaction among these components. First, we emphasize that it seems highly unlikely that the cosmological constant,  even though,  now occupies a  ``larger than life" status (with its present value of density parameter $\approx 0.73$) and also admittedly drives the acceleration in form of dark energy, and yet, it must lie dormant without falling into interaction with its close  counterpart (matter).

Second, it is worthwhile to recall here that the cosmological coincidence problem is hitherto one of the major unsolved issues in modern cosmology. Since their densities grow with evolution of universe  at different rates, there should be some fine-tuning  among the components. To solve  this issue,  several authors have attempted to  consider the possible interaction between dark energy and dark matter \cite{y3,y4,y5,y6,y7,y8,y9}. Recently, it has also been proposed that the interaction can sharply cut down the value of the cosmological constant during a middle phase of the universe, sandwiched between two non-interacting phases,  and thus solve the cosmological constant problem \cite{y1}.  As another dividend of such  decaying  cosmological constant, a mechanism has been suggested to generate dark matter \cite{y2} to siphon off  energy.  We study  below a  possible form of interaction between SCP and SDM,  and the consequent scaling of their energy densities as the universe evolves.

The laws of conservation of energy for SDM and SCP, allowing the interaction between them in phase II,   are given as
\begin{eqnarray}\dot{\rho'}_{m} + 3H(1+w'_{m})\rho'_{m}=Q\label{n12}\end{eqnarray}
\begin{eqnarray}\dot{\rho'}_{\lambda} + 3H(1+w'_{\lambda})\rho'_{\lambda}=-Q\label{n13}\end{eqnarray}  respectively,  where the interaction term $Q$  is  assumed to behave as
\begin{eqnarray}Q=\gamma\dot{\rho'}_{m}\label{n14}\end{eqnarray} with $\gamma$  as the constant of proportionality. Using (\ref{n14}) in (\ref{n12}) we obtain
\begin{eqnarray}\frac{\dot{\rho'}_{m}}{\rho'_{m}}=-\frac{3}{1-\gamma}\left(1-\frac{\varepsilon}{\dot{\phi}^{2}}+\varepsilon \right)\frac{\dot{a}}{a}\label{n15}.\end{eqnarray}

The present behaviour of the accelerating universe  appears  very similar  to inflation in  its early stage. Therefore,  we can take  $\dot{\phi}^{2}\approx 0$ (constant), just like the early inflationary era  which was dominated by cosmological constant (or slow-roll  of a scalar field with  $\dot{\phi}^{2}<<V(\phi)$) giving rise  to an exponential expansion.

 Solving  (\ref{n15}), we get
\begin{eqnarray}\rho'_{m}=\rho'^{0}_{m}\left(\frac{a_0}{a}\right)^\alpha\label{n16}\end{eqnarray}
with  \begin{eqnarray}\alpha=\frac{3}{1-\gamma}(1+\varepsilon-\varepsilon/\dot{\phi}^{2})\label{t1}.\end{eqnarray}
Here,  we see that as $\varepsilon\rightarrow 0$, $\rho'_{m}=\rho'^{0}_{m}\left(\frac{a_0}{a}\right)^{\frac{3}{1-\gamma}}$,  as may be  obtained in case of $w_{m}=0$ (interacting but normal dust matter).
In the  similar way,  we calculate $\rho'_{\lambda}$ from (\ref{n13}) by re-writing it as
\begin{eqnarray}\dot{\rho'}_{\lambda}=-\frac{\dot{a}}{a}\left[3\varepsilon\rho'_{\lambda}-\gamma\rho'^{0}_{m}\alpha\left(\frac{a_0}{a}\right)^{\alpha}\right]\label{n17}.\end{eqnarray}
Since  $w'_{m}+w'_{\lambda}=-1-\varepsilon/\dot{\phi}^{2}+2\varepsilon$, we have the inequality  $w'_{m}+w'_{\lambda}\neq w_{tach}$.  Thus,  $w'_{m}+w'_{\lambda}= w_{tach}- \delta$ with  $\delta=\varepsilon/\dot{\phi}^{2}+\dot{\phi}^{2}-2\varepsilon$.

For the SCP, solution of (\ref{n17}) is given as
 \begin{eqnarray}\rho'_{\lambda}=\rho'^{0}_{\lambda}x^{3\varepsilon} -\frac{\gamma\rho'^{0}_{m}(1-\varepsilon/\dot{\phi}^{2}+\varepsilon)}{1-\varepsilon/\dot{\phi}^{2}+\gamma\varepsilon} (x^\alpha-x^{3\varepsilon})\label{n18}\end{eqnarray}
where $x=a_{0}/a=1+z$,  (with $0\leq z\leq z_i$ in phase II) and,  particularly, in the absence of perturbation($\varepsilon\rightarrow 0$),  we have $\rho'_{\lambda}\rightarrow\rho'^{0}_{\lambda}-\gamma\rho'^{0}_{m}(x^{3/1-\gamma}-1)$.  This,  further, gives $\rho'_{\lambda}=\rho'^{0}_{\lambda}$ in the absence of interaction $(\gamma=0)$  as expected. The scaling behaviour  of  energy density of SCP indicates a possible solution to the cosmological constant problem stating as to why the present value of this constant is very small compared to that at the very early epoch following the big-bang.

The $(00)$  Friedmann equation in this phase, as  (\ref{m10}) in phase I,  for spatially flat universe ($k=0$) becomes

\begin{eqnarray}H^{2}(x)=H^{2}_{0}\left[\Omega'^{0}_{m}x^{\alpha}+\Omega'^{0}_{\lambda}x^{3\varepsilon}+\frac{\Omega'^{0}_{m}\gamma(1-\varepsilon/\dot{\phi}^{2}+\varepsilon)}{1-\varepsilon/\dot{\phi}^{2}+\gamma\varepsilon}\left(x^{3\varepsilon}-x^{\alpha}\right)\right]\label{m13}\end{eqnarray}
with $H_0$, $\Omega'^{0}_m$ and $\Omega'^{0}_{\lambda}$ refer to the present values of the Hubble parameter, and density parameters for matter  and cosmological constant respectively.
We can re-write  (\ref{n16}) and (\ref{n18}) in terms of redshift  and  determine the epoch of equality $z_{eq}$  at which $\rho'_{m}=\rho'_{\lambda}$ from
\begin{eqnarray}z_{eq}= \left[\frac{\Omega'^0_\lambda+ \beta\Omega'^0_m}{\Omega'^0_m (1+\beta)}\right]^\frac{1}{\alpha-3\varepsilon} -1\label{n19}\end{eqnarray} where \begin{eqnarray}\beta=\frac{\gamma(1-\varepsilon/\dot{\phi}^{2}+\varepsilon)}{1-\varepsilon/\dot{\phi}^{2}+\gamma\varepsilon}\label{qqq},\end{eqnarray}
  $\alpha$ is given by (\ref{t1}),  and $\Omega'^{0}_{\lambda}$ and  $\Omega'^{0}_{m}$ are the present values of density parameters  of cosmological constant and matter,  respectively.  Here, we take  $\Omega'^{0}_{\lambda}=0.73$ and $\Omega'^{0}_{m}=0.27$ \cite{i6}  or  $\Omega'^{0}_{\lambda}/\Omega'^{0}_{m}\approx 2.7037$ and the values  of $\gamma$ and $\varepsilon$ can be calculated from observations  discussed  next  in  Sec. 4.  In the absence of perturbation and interaction  both,  $z_{eq}\simeq0.3932$.
On the other hand,  if we consider a particular $z=z'$  when  the values  of $\dot{\rho'_{\lambda}}$ and $\dot{\rho'_{m}}$ are equal,  then we get
\begin{eqnarray}\frac{\rho'_{m}}{\rho'_{\lambda}}=\frac{\varepsilon (1-\gamma)}{(1+\gamma)(1-\varepsilon/\dot{\phi}^{2}+\varepsilon)}\label{n20}.\end{eqnarray} From (\ref{n20}),  we see that in the absence of perturbation $(\varepsilon =0)$ the ratio $\rho'_{m}/\rho'_{\lambda}=0$ which is un-physical. This implies that in  case of  unperturbed  EOS, the rates of fall of energy densities of these components cannot become equal.  Alternatively,  along with  $\gamma <1$,   both the components behave like SCP and SDM in this phase.

\section{\textbf{Diagnostics and calculation of interaction strength in  the  three-parameter space $(\gamma, \varepsilon,  \dot{\phi}^2)$ }}

Recently, Sahni \emph{et al} \cite{a6} introduced the redshift dependent function
\begin{eqnarray}O_{m}(x)=\frac{E^{2}(x)-1}{x^{3}-1}\label{n21}\end{eqnarray} where $E(x)=H(x)/H_{0}$ is the normalized Hubble function as derived   from (\ref{m13}). We attempt to find an approach to constrain the interaction strength between cosmological constant and matter by using $E(x)$.  Squaring it, we have \begin{eqnarray}E^{2}(x)=\Omega'^{0}_{m}x^{\alpha}+\Omega'^{0}_{\lambda}x^{3\varepsilon}+\frac{\Omega'^{0}_{m}\gamma(1-\varepsilon/\dot{\phi}^{2}+\varepsilon)}{1-\varepsilon/\dot{\phi}^{2}+\gamma\varepsilon}(x^{3\varepsilon}-x^{\alpha})\label{n22}\end{eqnarray} Using (\ref{n22}),  (\ref{n21}) becomes
 \begin{eqnarray}O_{m}(x)=\frac{(1-\gamma)(1-\varepsilon/\dot{\phi}^{2})\Omega'^{0}_{m}(x^{\alpha}-x^{3\varepsilon})+(1-\varepsilon/\dot{\phi}^{2}+\varepsilon)(x^{3\varepsilon}-1)}{(1-\varepsilon/\dot{\phi}^{2}+\varepsilon)(x^{3}-1)}\label{n23}\end{eqnarray}
It is clear that  as $\varepsilon\rightarrow 0$ and $\gamma\rightarrow 0$, $O_{m}(x)\rightarrow\Omega'^{0}_{m}$ which is just as expected.

From the  normalized Hubble function (\ref{n22}),  we calculate the difference $\Delta E^{2}(x_{i},x_{j})=E^2 (x_i) -E^2(x_j)$  for the pair of  two different redshifts  $x_{i}$ and $x_{j}$ as

\begin{eqnarray}\Delta E^{2}(x_{i},x_{j})=\Psi(x^{\alpha}_{i}-x^{\alpha}_{j})+\left(\Omega'^{0}_{\lambda}+\frac{\gamma\Omega'^{0}_{m}(1-\varepsilon/\dot{\phi}^{2}+\varepsilon)}{1-\varepsilon/\dot{\phi}^{2}+\gamma\varepsilon}\right)(x^{3\varepsilon}_{i}-x^{3\varepsilon}_{j})\label{n24}\end{eqnarray} where \begin{eqnarray}\Psi=\frac{\Omega'^{0}_{m}(1-\gamma)(1-\varepsilon/\dot{\phi}^{2})}{1-\varepsilon/\dot{\phi}^{2}+\gamma\varepsilon}.\label{m14}\end{eqnarray}

In case  $\varepsilon\rightarrow 0$,  we obtain
\begin{eqnarray}\Delta E^{2}(x_{i},x_{j})=(1-\gamma)\Omega'^{0}_{m}(x^{3/1-\gamma}_{i}-x^{3/1-\gamma}_{j}).\label{n26}\end{eqnarray}

For a given pair of redshifts,  the values of $\Delta E^{2}(x_{i},x_{j})$ can  be calculated using  data set of the Hubble parameter observations at   high- and low-$z$ epochs, both. The present value of the matter density parameter  $\Omega'^{0}_{m}$ is ascertained  from the concordance analysis of the Cosmic Microwave Background Radiation (CMBR) power spectrum  and  Baryon Acoustic Oscillation (BAO) data using the condition of spatial flatness.  The redshift pairs  used in these  calculations must,  however,  be sufficiently spaced over a wide range so that the  key-role of  $\gamma$,  and thus of $Q$ having the form (\ref{n14}),   in structure formation could be understood well. The reason is that  the  form and  strength of interaction between the components in phase II of the universe, mainly cosmological constant and matter, substantially influence the growth of structures, and thus may be easily constrained  by matching with the available structure surveys and power spectra  of the  WDM or $\Lambda$CDM  models   \emph{vis a vis} our approach adopted here. The data about the   precise measurements of $H_{0}$ will further  break the degeneracy in  the cosmological parameter  space\cite{i2}.

  Therefore,  with our choice  of  the   set of values of $H(z)$  at four epochs:  at $z=0.1$,  \quad $0.4$,   \quad $1.3$, and  \quad $1.75$,  the values of $H(z)=$ $69\pm12$,  \quad $95\pm17$,   \quad $168\pm17$ and   \quad $202\pm40$ km s$^{-1}$  Mpc$^{-1}$ respectively \cite{i3,i4}. The present value of Hubble parameter is used as  $H_{0}=73.8\pm2.4 \approx 73.8 $ km s$^{-1}$  Mpc$^{-1}$ \cite{i5} along  with  the present value of the density parameter of matter component  $\Omega'^{0}_{m}=0.272$  \cite{i6}. Thus,  correspondingly we have  $x_{1}=1.1$, \quad $x_{2}=1.4$, \quad $x_{3}=2.3$ and $x_{4}=2.75$  for these redshifts,  and the squared normalized  Hubble function  may be  calculated  as:

$E^{2}(x_{1}) = 0.874$, \quad $E^{2}(x_{2}) = 1.657$, \quad $E^{2}(x_{3})=5.182$,   \quad $E^{2}(x_{4})=7.492$  with their differences

 $\Delta E^{2}(x_{1},x_{2})=-0.783$, \quad $\Delta E^{2}(x_{1},x_{3}) =-4.308$,

\quad $\Delta E^{2}(x_{1},x_{4})=-6.618$,\quad $\Delta E^{2}(x_{2},x_{3})=-3.525$,

\quad $\Delta E^{2}(x_{2},x_{4})=-5.835$ and $\Delta E^{2}(x_{3},x_{4})=-2.310$

 These data lead  to  determine six values of the proportionality constant $\gamma$ through following expressions\\

$\Psi\left[(1.1)^{\alpha}-(1.4)^{\alpha}\right]+(1-\Psi)[(1.1)^{3\varepsilon}-(1.4)^{3\varepsilon}] = -0.783$,\\

$\Psi\left[(1.1)^{\alpha}-(2.30)^{\alpha}\right]+(1-\Psi)[(1.1)^{3\varepsilon}-(2.30)^{3\varepsilon}]= -4.308$,\\

$\Psi\left[(1.1)^{\alpha}-(2.75)^{\alpha}\right]+(1-\Psi)[(1.1)^{3\varepsilon}-(2.75)^{3\varepsilon}]=  -6.618$,\\

$\Psi\left[(1.4)^{\alpha}-(2.30)^{\alpha}\right]+(1-\Psi)[(1.4)^{3\varepsilon}-(2.30)^{3\varepsilon}]= -3.525$,\\

$\Psi\left[(1.4)^{\alpha}-(2.75)^{\alpha}\right]+(1-\Psi)[(1.4)^{3\varepsilon}-(2.75)^{3\varepsilon}]= -5.835$,\\

$\Psi\left[(2.30)^{\alpha}-(2.75)^{\alpha}\right]+(1-\Psi)[(2.30)^{3\varepsilon}-(2.75)^{3\varepsilon}]= -2.310$. \\

The solution of the above set can be attained by further imposing viable conditions to break degeneracy in the three-parameter space $(\gamma, \varepsilon,   \dot{\phi}^2)$ in order to extract information about the single parameter $\gamma$.

\section{\textbf{Role of interaction in cosmic age} }
The age of universe at a redshift $z$ can be calculated by integrating
\begin{eqnarray} dt=-\frac{dz}{(1+z)H(z)}\label{n27}\end{eqnarray}
between the corresponding limits of redshifts $[z,\infty]$. In a two-phase model discussed in this paper, the total  age of universe at the present epoch  is the sum of duration of  the  non-interacting phase and that of the  interacting one.  Therefore, we may  find the time duration of universe $t_i$ in phase I,   from (\ref{m10}) and (\ref{n27}) as
\begin{eqnarray}t_{i}=\int^{\infty}_{z_{i}}\frac{dz}{(1+z)H_{i}\left[\Omega^{i}_{m}\left(\frac{1+z}{1+z_{i}}\right)^{3}+\Omega^{i}_{r}\left(\frac{1+z}{1+z_{i}}\right)^{4}+\Omega^{i}_{\lambda}\right]^{1/2}}\label{m15}\end{eqnarray}
 and in phase II,  from (\ref{m13}) (with the total age at present  taken to be  $t_0$),  as
\begin{eqnarray}t_{0}-t_{i}=\int^{z_{i}}_{0}\frac{dz}{(1+z)H_{0}\left[\Omega'^{0}_{m}(1+z)^{\alpha}+\Omega'^{0}_{m}\beta\{(1+z)^{3\varepsilon}-(1+z)^{\alpha}\}+\Omega'^{0}_{\lambda}(1+z)^{3\varepsilon}\right]^{1/2}}\label{m16}\end{eqnarray}

Thus,  the  total  age of the universe at the present epoch depends on three parameters,  $\gamma$, $\varepsilon$ and $\dot{\phi}^{2}$, that is,  $t_{0}=f(\gamma,\varepsilon,\dot{\phi}^2)$.  The  parameters $\gamma$ and $\varepsilon$ can be obtained from observations,  while $\dot{\phi}^{2}$ is not directly fixed by  them.  Clearly,  the age is directly   influenced by  interaction, and with the unperturbed EOS of the components in this phase,  it is given as

\begin{eqnarray}t_{0}=\int^{\infty}_{z_{i}}\frac{dz}{(1+z)H_{i}\left[\Omega^{i}_{m}\left(\frac{1+z}{1+z_{i}}\right)^{3}+\Omega^{i}_{r}\left(\frac{1+z}{1+z_{i}}\right)^{4}+\Omega^{i}_{\lambda}\right]^{1/2}} \nonumber \\ +\int^{z_{i}}_{0}\frac{dz}{(1+z)H_{0}\left[(1-\gamma)\Omega'^{0}_{m}(1+z)^{3/1-\gamma}+\gamma\Omega'^{0}_{m}+\Omega'^{0}_{\lambda}\right]^{1/2}}\label{m17}\end{eqnarray}

\section{\textbf{Conclusion }}

In this paper, we have resolved  the tachyonic scalar  field into its possible components, and studied their  evolution in the expanding universe spanned over two phases, that is, the  non-interacting phase  and the interacting one. In the first phase I,  the field breaks into two components, namely radiation and the shifted cosmological parameter (SCP). The SCP is further found to consist of two components---true cosmological constant and some form of exotic pressure-less  matter with negative energy density. In this phase, these three components do not interact mutually,  and,  therefore in Sec.  2,  we find their specific evolution with respect to the redshift of the large scale cosmic expansion.

In the following phase II, as examined in  Sec. 3,  radiation  component descending from phase I is suppressed, and the tachyonic field is dominated by two components---matter and true cosmological constant. We introduce a small constant perturbation  $\varepsilon$ (indeed, it may be time-dependent, in general, and we would take up such form elsewhere) into the EOS of cosmological constant, which in turn changes the EOS of the matter component as well. These components are then,  respectively,  known as the shifted cosmological parameter (SCP) and  the shifted dust matter (SDM) of phase II,  and we allow them to interact by  choosing a specific form (\ref{n14}),  so as to determine the role of this interaction on their cosmic  evolution, and in turn on the dynamics of the universe.  We also determine the redshift  $z_{eq}$ when the SCP and SDM have equal energy densities and obtain a condition that applies when the corresponding rates of their scaling become equal. This highlights  mainly the role of $\gamma$ and $\varepsilon$ in determining the evolution of dark energy in our  model.

In Sec. 4, we attempted  to frame a normalized  Hubble  function, earlier used in $O_m(x)$ diagnostic, to constrain the interaction between the components of phase II.  The low- and high-redshift observations for Hubble parameter and the concordance values of the density parameters from CMBR, BAO data-sets lead us to construct equations for determining the strength of interaction. However, in the three-parameter space $(\gamma,  \varepsilon,  \dot{\phi}^2)$, further conditions must be imposed to extract information about interaction alone.  We will attempt to devise such reasonable  conditions  to this end   in future. We also emphasize that the interaction  substantially modifies  the growth of structures in past,  and so,  the observations  must be included over a sufficiently  large range of redshifts in the post-recombination era,  co-extensive with the structure formation,  to enable us to examine the distinguishing imprints on the power spectrum.  With this view,  in a future study, our  model may  be further developed and compared with the detailed analysis of the  power spectra  of  WDM or $\Lambda$CDM   models.

Finally, we proceeded  to  find in Sec. 5   the influence of interaction in  phase II in determining  the present age of the universe. This was  calculated by including the time  spans of both phases.  Again, re-affirming our findings  in Sec. 4, the  total  cosmic age too exhibits a functional form  in the three-parameter space,  only to be constrained with appropriate observational and/or  \emph{a priori}  conditions.  We would endeavour  to address these issues in our future work.

\section{\textbf{Acknowledgments} } The authors thankfully acknowledge the financial support F. No. 37-431/2009 (SR) received from UGC, New Delhi regarding this work.

\end{document}